\documentclass[sigconf]{acmart}

\usepackage{graphicx}
\usepackage{amsmath}
\usepackage{booktabs}

\title{Improving Visual Recommendation on E-commerce Platforms Using Vision-Language Models}

\author{Yuki Yada}
\email{y-b-yada@mercari.com}
\affiliation{%
  \institution{Mercari, Inc.}
  \country{}
}

\author{Sho Akiyama}
\email{s-akiyama@mercari.com}
\affiliation{%
  \institution{Mercari, Inc.}
  \country{}
}

\author{Ryo Watanabe}
\email{r-b-watanabe@mercari.com}
\affiliation{%
  \institution{Mercari, Inc.}
  \country{}
}

\author{Yuta Ueno}
\email{y-b-ueno@mercari.com}
\affiliation{%
  \institution{Mercari, Inc.}
  \country{}
}

\author{Yusuke Shido}
\email{shidoy@mercari.com}
\affiliation{%
  \institution{Mercari, Inc.}
  \country{}
}

\author{Andre Rusli}
\email{andre.rusli@mercari.com}
\affiliation{%
  \institution{Mercari, Inc.}
  \country{}
}

\copyrightyear{2025}
\acmYear{2025}
\setcopyright{rightsretained}
\acmConference[RecSys '25]{Proceedings of the Nineteenth ACM Conference on Recommender Systems}{September 22--26, 2025}{Prague, Czech Republic}
\acmBooktitle{Proceedings of the Nineteenth ACM Conference on Recommender Systems (RecSys '25), September 22--26, 2025, Prague, Czech Republic}
\acmDOI{10.1145/3705328.3748128}
\acmISBN{979-8-4007-1364-4/2025/09}

\begin{document}

\begin{abstract}
On large-scale e-commerce platforms with tens of millions of active monthly users, recommending visually similar products is essential for enabling users to efficiently discover items that align with their preferences. This study presents the application of a vision-language model (VLM)---which has demonstrated strong performance in image recognition and image-text retrieval tasks---to product recommendations on Mercari, a major consumer-to-consumer marketplace used by more than 20 million monthly users in Japan. Specifically, we fine-tuned SigLIP, a VLM employing a sigmoid-based contrastive loss, using one million product image-title pairs from Mercari collected over a three-month period, and developed an image encoder for generating item embeddings used in the recommendation system. Our evaluation comprised an offline analysis of historical interaction logs and an online A/B test in a production environment. In offline analysis, the model achieved a 9.1\% improvement in nDCG@5 compared with the baseline. In the online A/B test, the click-through rate improved by 50\% whereas the conversion rate improved by 14\% compared with the existing model. These results demonstrate the effectiveness of VLM-based encoders for e-commerce product recommendations and provide practical insights into the development of visual similarity-based recommendation systems.

\end{abstract}


\maketitle

\section{Introduction}

As millions of new items are listed daily on e-commerce platforms, users encounter growing challenges in locating products that match their preferences.

Visual similarity-based recommendations that leverage features such as color, shape, and patterns that are not captured in text data have become essential. Platforms such as Pinterest \cite{zhai2019pinterest}, Amazon \cite{du2022amazon}, and eBay \cite{yang2017ebay} have implemented these systems. In particular,  this is important in consumer-to-consumer (C2C) marketplaces such as Mercari \footnote{https://jp.mercari.com/}, where user-generated listings comprise a diverse range of unique, often second-hand items that lack standard product identifiers and consistent textual descriptions. In these contexts, visual cues play a critical role in bridging the information gap and facilitating effective discovery.

\begin{figure}[t]
\centering
\includegraphics[width=0.85\linewidth, alt={Overview diagram of a visual product recommendation system. A product image from a mobile app is processed by an Image Encoder to generate an image embedding. This embedding is used to query a Vector Store, which retrieves visually similar products that are then displayed in the app, typically below the original product.}]{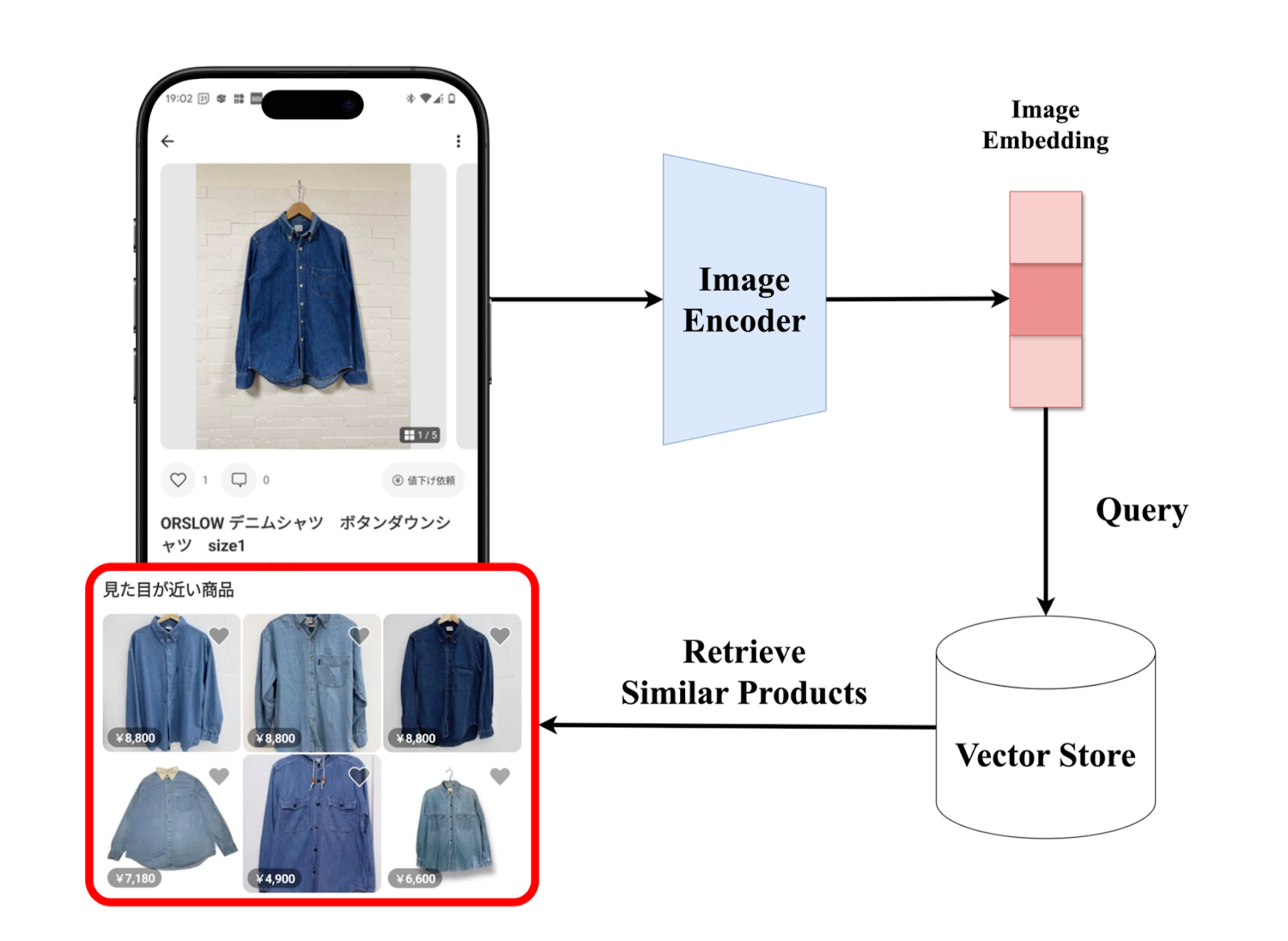}
\caption{Overview of a product recommendation system based on visual similarity.}
\label{fig:overview}
\end{figure}

A typical pipeline for visual similarity-based recommendations, depicted in Figure \ref{fig:overview}, generally involves the following steps:

\begin{enumerate}
    \item Convert the query product image into a vector representation.
    \item Perform nearest-neighbor search in a database of image vectors to retrieve similar products.
\end{enumerate}

The performance of these systems depends significantly on the image encoder quality. Although efficient models such as ResNet \cite{he2016resnet} and MobileNet \cite{howard2017mobilenets, mark2018mobilenetv2} are widely used \cite{zhai2019pinterest, du2022amazon}, they often fail to capture fine-grained features and cross-category similarities.

Recently, vision-language models (VLMs) trained on large-scale image-text pairs have outperformed conventional models across multiple benchmarks \cite{zhai2023sigmoid, radford2021clip}. The present study investigated the effectiveness of a VLM-based image encoder for product recommendations on Mercari, a leading C2C marketplace with over 20 million monthly active users. The key contributions of this study are as follows:

\begin{enumerate}
    \item We demonstrate, via offline evaluation, that the proposed VLM-based image encoder significantly outperforms the baseline model, which employs a traditional convolutional neural network (CNN) encoder pre-trained for image classification tasks, achieving a 9.1\% improvement in nDCG@5.
    \item We successfully deployed the model in a production environment on a large-scale e-commerce platform and verified its effectiveness via live A/B testing, resulting in a 50\% increase in click-through rate (CTR) and 14\% increase in conversion rate (CVR) for visually similar product recommendations.
\end{enumerate}

\section{Related Work}

\subsection{Vision-Language Models}

CLIP \cite{radford2021clip} achieved notable results via multimodal and zero-shot learning. Although not directly optimized for specific benchmarks, it exhibited strong performance across a range of tasks, including image classification \cite{conde2021clipart} and video retrieval \cite{fang2021clip2video}. Subsequently, the SigLIP model \cite{zhai2023sigmoid} addressed the limitations of prior methods, including CLIP---specifically, the overreliance of conventional softmax-based contrastive losses on in-batch negative samples. SigLIP addresses this issue by employing a sigmoid loss function. In image-text retrieval tasks on standard datasets such as MS COCO \cite{chen2015coco} and Flickr30k \cite{plummer2016flickr30k}, SigLIP demonstrated superior performance compared with conventional softmax-based contrastive losses, particularly in zero-shot and transfer learning scenarios. Furthermore, its stability across different batch sizes and learning rates was experimentally validated \cite{zhai2023sigmoid}. Therefore, in this study, SigLIP was used as the core model for image retrieval.

\subsection{Real-world Applications}

In recent years, image recognition technologies have gained traction across a range of industries. Systems developed by companies such as Meta \cite{tang2019msuru}, Pinterest \cite{zhai2019pinterest}, and Google \cite{googlelens} are deployed in production environments and contribute to improvements in sales and customer experience. Visual search has become a standard component of e-commerce platforms (\cite{du2022amazon}, \cite{yang2017ebay}). However, various existing visual search systems are built from scratch and predominantly focus on localizing relevant items within query images and retrieving category-level similar items. In contrast, relatively few studies have explored the direct application of advanced image recognition or multimodal models to generate business outcomes. This study demonstrated the improvement of business KPIs with a relatively simple configuration employing a fine-tuned SigLIP model, which outperformed an existing baseline image recognition model.

\section{Visual Recommendation Using SigLIP}\label{sec:method}

\begin{figure}[t]
\centering
\includegraphics[width=0.85\linewidth, alt={A training pipeline diagram for contrastive pre-training of text and image pairs. Text inputs are processed by a Text Encoder and image inputs by an Image Encoder to generate their respective embedding vectors. A similarity matrix on the right displays pairwise scores between these text and image embeddings, with scores for matched pairs highlighted along the diagonal.}]{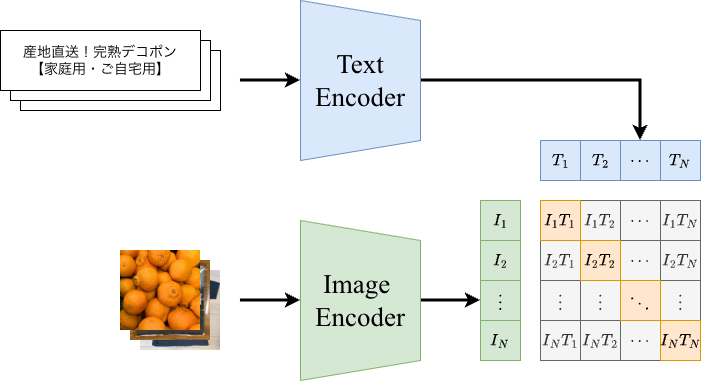}
\caption{Training pipeline. $I_1, \ldots, I_n$ represent the embedding vectors of product images 1 to $N$, and $T_1, \ldots, T_n$ represent the corresponding embedding vectors of product titles 1 to $N$.}
\label{fig:training}
\end{figure}

We developed a visual similarity-based recommendation system for Mercari. The SigLIP model, pre-trained on the WebLI dataset \cite{chen2022webli}, was fine-tuned using product image-title pairs from Mercari listings over a three-month period. The image encoder uses the ViT B/16 architecture, whereas the text encoder is a B-sized transformer. Figure \ref{fig:training} illustrates the training pipeline, wherein each image-title pair is encoded and trained employing contrastive loss.

Using a fine-tuned image encoder, we generated vector embeddings for Mercari product images and indexed them into a vector store containing tens of millions of items. For recommendations, given an image embedding of a query product, we retrieved visually similar items by performing an approximate nearest neighbor (ANN) search over the indexed embeddings within this store.

\section{Evaluation}\label{sec:eval}

\subsection{Model Training}\label{subsec:model_train}

We utilized real product data from April 29 to July 29, 2024, comprising approximately 853 million listings. After excluding ``reserved’’ listings (items designated for specific buyers, where titles frequently function as direct messages rather than accurate image descriptions, making them unsuitable for our image-title pair training), we sampled one million image-title pairs.

We fine-tuned the multilingual SigLIP model (\textit{google/siglip-base-patch16-256-multilingual}) employing contrastive learning. Training was performed for 5 epochs with a batch size of 256 and learning rate of 5e-5 on NVIDIA L4 GPUs.

\subsection{Offline Evaluation}\label{subsec:offline_eval}

We conducted an offline evaluation to compare the performance of our fine-tuned SigLIP model against the ImageNet \cite{jia2009imagenet} pretrained MobileNetV2 \cite{mark2018mobilenetv2} (\textit{google/mobilenet\_v2\_1.4\_224}) image encoder previously used in the production environment of Mercari. This evaluation used historical user interaction logs, specifically user impressions and taps, to measure the relevance of visually similar product recommendations. We employed standard information retrieval metrics, including nDCG@k and precision@k.

\begin{table}[t]
\centering
\caption{SigLIP achieved a 9.1\% gain in nDCG@5 and 15.7\% in Precision@1 over MobileNetV2.}
\begin{tabular}{lccc}
\toprule
Model & nDCG@5 & Precision@1 & Precision@3 \\
\midrule
MobileNetV2 & 0.607 & 0.356 & 0.601 \\
SigLIP & \textbf{0.662} & \textbf{0.412} & \textbf{0.660} \\
SigLIP + PCA & 0.647 & 0.406 & 0.658 \\
\bottomrule
\end{tabular}
\label{tab:offline_results}
\end{table}

The results summarized in Table \ref{tab:offline_results} demonstrate the superiority of the VLM-based approach. The fine-tuned SigLIP model achieved an nDCG@5 score of 0.662, representing a significant 9.1\% improvement over the MobileNetV2 baseline score of 0.607. Furthermore, precision@1 demonstrated a significant improvement of 15.7\%, increasing from 0.356 for MobileNetV2 to 0.412 for SigLIP. These findings indicate that the VLM-based image encoder is capable of extracting more pertinent visual features from e-commerce product images than the conventional CNN-based model.

To improve deployment efficiency, we reduced the SigLIP embedding dimension from 768 to 128 using PCA fit on 20 million product embeddings. This strategy significantly reduces vector storage requirements (by approximately 83\%) with minimal impact on recommendation quality. As detailed in Table \ref{tab:offline_results}, the resulting SigLIP + PCA model exhibited only a slight decrease in nDCG@5 (-2.3\% compared to the full SigLIP, reaching 0.647) while still maintaining a substantial 6.6\% lead over the MobileNetV2 baseline. This confirms that PCA provides an effective trade-off, enabling considerable resource savings while largely preserving the accuracy gains achieved by the fine-tuned VLM.

\subsection{Model Deployment}\label{sec:model_deployment}

\begin{figure}[t]
\centering
\includegraphics[width=0.85\linewidth, alt={System architecture diagram illustrating an inference pipeline with two main flows. The first flow, labeled 'Async embedding worker,' describes how new listing thumbnails are processed by an Image Encoder to generate an item ID and an embedding vector, which are then stored in a Vector Store. The second flow shows user interaction: when a user views an item on an app, the item's ID initiates an Approximate Nearest Neighbor (ANN) search against the Vector Store. The search results subsequently pass through Filtering and Re-ranking stages to produce a final list of recommended items like Item A, Item B, etc.}]{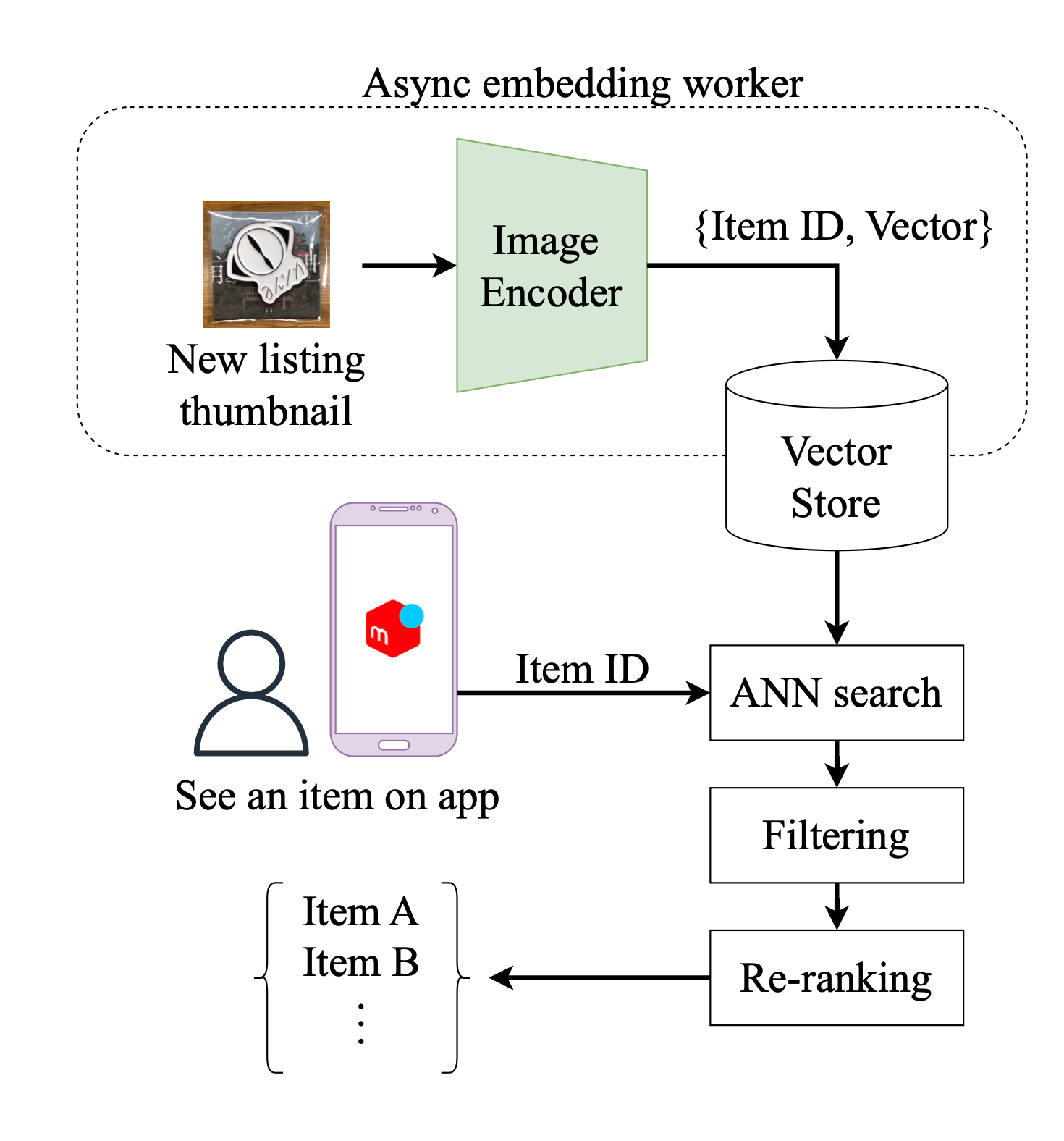}
\caption{System Architecture}
\label{fig:inference}
\end{figure}

The production system architecture, illustrated in Figure \ref{fig:inference}, employs an asynchronous pipeline for embedding preparation and a real-time service for recommendation generation.

Upon listing a new item, the asynchronous pipeline generates and indexes embeddings. It utilizes the fine-tuned SigLIP image encoder and PCA transformation (detailed in Sections \ref{sec:method} and \ref{subsec:offline_eval}) to compute a 128-dimensional embedding for each product image. These embeddings, in conjunction with the item IDs, are stored in a vector store.

When a user views an item, the real-time service retrieves the pre-computed 128-dimensional embedding for that item from the vector store. This embedding acts as a query for an ANN search performed against the indexed data. This search efficiently returns a list of candidate item IDs deemed visually similar to the query item.

As depicted in Figure \ref{fig:inference}, these retrieved candidates subsequently undergo filtering and re-ranking stages to produce the final list presented to the user. Specifically:

\begin{itemize}
    \item Filtering: Candidate items are filtered based on predefined rules; those with prices significantly deviating from the query item are removed.
    \item Re-ranking: The remaining candidates are subsequently re-ranked based on their category similarity to the query item, aiming to enhance the final perceived relevance.
\end{itemize}

\subsection{Online A/B Test}

To evaluate the real-world performance of the proposed model, we conducted an online A/B test on the Mercari platform, specifically targeting the ``Visually Similar Items’’ section on product detail pages. The experiment compared the recommendations generated using embeddings from our fine-tuned SigLIP model (with PCA applied, yielding 128 dimensions) as the treatment group with that of the conventional MobileNetV2-based embeddings that served as the control group.

The results demonstrated significant improvements in the SigLIP-based approach. Compared with the control group, the treatment group exhibited a 50\% increase in the CTR for the recommended items and 14\% increase in the CVR (purchases originating from clicks on these recommendations). These findings from the live production environment confirm that the fine-tuned SigLIP encoder delivers substantial gains in user engagement and discovery effectiveness compared with the baseline model.

\section{Conclusion}
We introduced a visual similarity-based product recommendation system using a fine-tuned SigLIP on a large-scale e-commerce platform. 

The obtained results demonstrate:

\begin{itemize}
    \item \textbf{9.1\% improvement} in offline evaluation metrics.
    \item \textbf{50\% increase} in CTR and \textbf{14\% increase} in purchase conversions in live testing.
\end{itemize}

These findings confirm that VLM-based image encoders are highly effective for product recommendations in e-commerce. Future studies will include integrating multimodal data and developing personalized recommendation systems based on user preferences.

% \section{Acknowledgement}

% Thank you for paul

\section{Authors Bio}

Yuki Yada, Sho Akiyama, Ryo Watanabe, Yuta Ueno, Andre Rusli, and Yusuke Shido are Machine Learning Engineers at Mercari, Inc. Mercari is one of Asia's leading C2C marketplaces, serving more than 20 million active monthly users. Yuki Yada, Sho Akiyama, and Andre Rusli focused on product improvements and research, leveraging generative AI and large language models within Mercari. Ryo Watanabe, Yuta Ueno, and Yusuke Shido concentrated on the research and enhancement of recommendation systems for the platform.

\section{Acknowledgments}

We would like to express our sincere gratitude to Shinya Yaginuma, Product Manager of the Recommendation ML team at Mercari, for his invaluable leadership and dedication in driving this visual recommendation project forward. His strategic vision and continuous support were instrumental in bringing this research from conception to successful deployment in our production environment.

\bibliographystyle{ACM-Reference-Format}
\bibliography{base}
% \begin{thebibliography}{99}

% \end{thebibliography}

\end{document}